\documentstyle[11pt]{article}
\newcommand{\arccot}{{\rm arccot}}

\begin{document}
\large

 {\LARGE{\bf
 \centerline {Global structure of Robinson-Trautman}
 \centerline {radiative space-times with cosmological constant}}}
 \vspace{20mm}
 \centerline {Ji\v r\'\i\  Bi\v c\' ak    $^{1,2}$ and
              Ji\v r\' \i\  Podolsk\' y   $^1$}
 \vspace{15mm}
 {\it
 \centerline {$^1$ Department of Theoretical Physics,}
 \centerline {Faculty of Mathematics and Physics, Charles University,}
 \centerline {V Hole\v sovi\v ck\'ach 2, 180 00 Prague 8, Czech Republic}
 \vspace{10mm}
 \centerline {$^2$ Max-Planck-Institute for Gravitational Physics,}
 \centerline {Schlaatzweg 1, Potsdam, D-14473, Germany}}
 \vspace{10mm}
 {\footnotesize
 \centerline {Electronic addresses:
              bicak@mbox.troja.mff.cuni.cz,
              podolsky@mbox.troja.mff.cuni.cz}}

 \vspace{10mm}
\begin{abstract}
Robinson-Trautman radiative space-times of Petrov type II with a
non-vanishing cosmological constant $\Lambda$ and mass
parameter $m>0$ are studied using analytical
methods. They are shown to approach the corresponding spherically symmetric
Schwarzschild-de Sitter or Schwarzschild-anti-de Sitter solution
at large retarded times. Their global structure is analyzed, and it is
demonstrated that the smoothness of the extension of the metrics across
the horizon, as compared with the case  $\Lambda=0$,
is increased for $\Lambda>0$ and decreased for $\Lambda<0$.
For the extreme value  $9\Lambda m^2=1$, the extension is smooth but
not analytic. This case appears to be the first example of a
smooth but not analytic horizon.
The models with  $\Lambda>0$ exhibit explicitly the
cosmic no-hair conjecture  under the presence of gravitational waves.

\vspace{3mm}
\noindent
PACS number(s): 04.30.-w, 04.20.Jb, 98.80.Hw
\end{abstract}

\newpage
\section {Introduction}

   Robinson-Trautman vacuum space-times \cite{RT1, RT2}
have attracted increased attention in the last decade,
in particular in the works by
Luk\'acs et al. \cite{Luk}, Schmidt \cite{Schm},
Rendall \cite{Ren} and, most recently, by Chru\'sciel and Singleton
\cite{Sin}-\cite{ChruSin}. (We refer the reader to the last papers for
further references.)  In these studies the Robinson-Trautman
space-times were shown to exist globally for all positive
``times'', and to converge asymptotically to a Schwarzschild
metric. This global time behaviour is true for generic, arbitrarily
strong smooth initial data within the class of the Robinson-Trautman
space-times. Interestingly, the extension of these space-times across
the  ``Schwarzschild-like'' event horizon can only be made with a
finite degree of smoothness.

   The Robinson-Trautman metrics can easily be generalized to
solve the vacuum Einstein equations with a non-vanishing
$\Lambda$ \cite{KSMH}. The results proving the global existence
and convergence of the  Robinson-Trautman solutions
can be taken over from previous studies since
$\Lambda$ does not explicitly enter the basic Robinson-Trautman
equation. However, the presence of $\Lambda$ has a considerable
effect on the global structure of the space-times. In our
previous work \cite{BiPo}, we demonstrated that the
Robinson-Trautman  space-times of the Petrov type II with
$\Lambda>0$ such that $9\Lambda m^2<1$ settle down to the
Schwarzschild-de Sitter space-time at large retarded times. They
admit a smooth future space-like infinity and continuation of the
metric across the ``Schwarzschild-de Sitter-like'' black-hole horizon
can be made with a  higher degree of smoothness than in the corresponding
cases with $\Lambda=0$. These space-times may serve as exact models
of black-hole formation in non-spherical space-times which are not
asymptotically flat. They also represent the
only known exact analytic demonstration of the cosmic no-hair
conjecture (see, e.g., \cite{Gibb}-\cite{Barrow})
under the presence of gravitational waves.

The analysis in \cite{BiPo}, however, covers only the cases with
$\Lambda$ and $m$ such that $0<9\Lambda m^2<1$, implying the existence
of both the black-hole and  cosmological horizons.
The purpose of this work is to study the `extreme' case
with $9\Lambda m^2=1$, in which the two horizons coincide, and
the cases with $9\Lambda m^2>1$, when the naked singularity
arises. We also analyze the global structure of the Robinson-Trautman
space-times with $\Lambda<0$, which admit one black-hole horizon.

The formation of an extreme Reissner-Nordstr\"om black hole in
collapse with small nonspherical perturbations
\cite{Bi72, Bic77b}, as well as motion of particles in
extreme black-hole space-times \cite{BS} exhibit features qualitatively
different from those of generic black holes.
Perturbations of extreme black holes seem to be
stable with respect to both classical  and quantum processes, and
there are attempts to interpret them as solitons
\cite{Haj, Gibb91}. Extreme black holes with cosmological
constant were discussed by Lake and Roeder \cite{Lake},
Mellor and Moss \cite{MeMo1, MeMo2}, Romans \cite{Romans},
Brill and Hayward \cite{BH}, and others.
They were also studied in the context of the
Einstein-Yang-Mills-Higgs theory (see, e.g., \cite{GHT, BCHH} and
references therein).

Very recently, Kastor and Traschen \cite{KaTr} have given the
solutions with a cosmological constant $\Lambda>0$, containing many
extreme black holes. The solutions were used for
analytic studies of black-hole collisions
and cosmic censorship hypothesis \cite{BHKT}. Horizons of
these space-times were analyzed in detail in
\cite{BHKT}-\cite{Welch}.

It is noteworthy that multi-black-hole solutions consisting of
the analogues of extremal Reissner-Nordstr\"om black holes in
asymptotically de Sitter space-time have horizons that are not
smooth \cite{BHKT}. In contrast to such black holes in
asymptotically flat space-times which have smooth horizons and
are static, the cosmological multi-black-hole solutions are
dynamic with gravitational and electromagnetic radiation. The fact
that horizons are not smooth is interpreted as due to the
existence of radiation which does not have a ``smooth
distribution''.  It would then seem natural to interpret the
non-smoothness of the horizons of the Robinson-Trautman black
holes in a similar way. On the other hand, one should bear in
mind that in five or more dimensions some multi-black-hole
solutions to $d$-dimensional Einstein gravity have horizons that
are not smooth although these solutions are static \cite{Welch}.
Their lack of smoothness thus cannot be attributed to the presence
of radiation.

   In the  next section we briefly summarize results
of recent studies of the Robinson-Trautman vacuum space-times
with $\Lambda=0$, and, in Section 3, we review results for
the Robinson-Trautman space-times  with $0< 9\Lambda m^2<1$,
including their global structure and asymptotic properties
at future infinity. Section 4 and Section 5 are devoted to
analysis of the `extreme' case ($9\Lambda m^2=1$) and
`naked-singularity' cases ($9\Lambda m^2>1$). In Section 6 the
Robinson-Trautman space-times  with $\Lambda<0$ are studied.
The results are summarized and some general remarks added in the
concluding Section 7.

\section {The Robinson-Trautman space-times with $\Lambda=0$}

   In the standard form the Robinson-Trautman vacuum metric reads
(see \cite{RT1,RT2, KSMH})
\begin{equation}
ds^2=-\Phi du^2 - 2dudr + 2r^2P^{-2}d\zeta d\bar\zeta\ , \label{E2.2}
\end{equation}
where $P=P(u,\zeta,\bar\zeta)$, $\zeta$ is a complex spatial
coordinate, $r\in[0,\infty)$ is the affine
parameter along the rays $u=const$, $\zeta=const$, and
\begin{equation}
\Phi=\Delta\ln P -2r(\ln P)_{,u} - \frac{2m}{r}\ .        \label{E2.3}
\end{equation}
Here $\Delta=2P^2\partial^2/\partial\zeta\partial\bar\zeta$
and $m$ is a constant related to the Bondi mass of the system.
The function $P$ satisfies the Robinson-Trautman equation
\begin{equation}
(\ln P)_{,u}=-\frac{1}{12m}\Delta\Delta(\ln P)\ .          \label{E2.4}
\end{equation}
This equation can be formulated (see, e.g.,
\cite{Chru1, Chru2, ChruSin})
by introducing a smooth metric $g^{0}_{ab}(x^c)$ on a
two-dimensional manifold
(here we shall concentrate on the physical case $S^2$)
and a $u$-dependent family of 2-metrics
$g_{ab}=\left[ f(u,x^c)\right]^{-2}\, g^{0}_{ab}\,$
which, with respect to the  coordinate $ \zeta$, take the form
$2P^{-2} d\zeta d\bar\zeta$. Writing
\begin{equation}
P=f\,P_0 \ ,\qquad P_0=1+\frac{1}{2}\zeta\bar\zeta\ ,     \label{E2.10}
\end{equation}
we find equation (\ref{E2.4}) becomes
\begin{equation}
{\partial f\over\partial u}= - \frac{f}{24m}\Delta_g R\ ,    \label{E2.6}
\end{equation}
where $R$ is the curvature scalar and $\Delta_g$ the Laplacian of
the metric $g_{ab}$. Using $R_0$ and $\Delta_0$ to denote the curvature
scalar and the Laplacian of $g^{0}_{ab}$, one has
\begin{equation}
R=f^2(R_0+2\Delta_0\ln f),\ \ \ \Delta_g=f^2\Delta_0 \ .      \label{E2.7}
\end{equation}
Choosing standard coordinates on the sphere,
$\zeta=\sqrt2 e^{i\varphi}\tan{\theta/2}$, we obtain
\begin{equation}
2P_0^{-2}d\zeta d\bar\zeta=d\theta^2+\sin^2\theta d^2\varphi \ ,
\qquad \Delta_0\ln P_0=1\ ,\qquad R_0=+2\ .\label{E2.9}
\end{equation}
Therefore, the metric (\ref{E2.2}) with $P=P_0$ is just the Schwarzschild
metric,
\begin{eqnarray}
ds^2&=&-\left(1-\frac{2m}{r}\right)du^2-
  2dudr+r^2(d\theta^2+\sin^2\theta d\varphi^2) \nonumber \\
    &=&-\left(1-\frac{2m}{r}\right)dt^2+
      \left(1-\frac{2m}{r}\right)^{-1}dr^2+
      r^2(d\theta^2+\sin^2\theta d\varphi^2) \ ,   \label{E2.9a}
\end{eqnarray}
where $u=t-r^*$, and $r^*=\int \Phi^{-1}(r)\,dr=r+2m\ln(r/2m-1)$
is the usual ``tortoise'' coordinate.

   The most general analysis of the existence and behavior
of solutions of the Robinson-Trautman equation was recently given
by Chru\'sciel \cite{Chru1, Chru2} and by Chru\'sciel and Singleton
\cite{ChruSin} (cf. also \cite{Schm, Sin, Tod}). The main result is
that when $f_0\equiv f(u=u_0,x^a)$ is an arbitrary, sufficiently smooth
initial-value function for $f$, then $f$ satisfying equations (\ref{E2.6})
and (\ref{E2.7}) exists for all times $u\ge u_0$;
an asymptotic expansion of $f(u,x^a)$ for large $u$ has the form
\begin{eqnarray}
f= \sum_{i,j\ge0} f_{i,j} u^j e^{-2iu/m}=
1+f_{1,0}e^{-2u/m}+f_{2,0}e^{-4u/m}+\cdots+f_{14,0}e^{-28u/m} \nonumber\\
   +f_{15,1}ue^{-30u/m}+f_{15,0}e^{-30u/m}+\cdots \ ,   \label{E2.13}
\end{eqnarray}
where $f_{i,j}$ are smooth functions on
$S^2$. Therefore, as $u\rightarrow+\infty$, {\it Robinson-Trautman
metrics approach exponentially fast a Schwarzschild metric},
$f=1$. (In general $f\rightarrow f_{Schw}$, where $f_{Schw}$
corresponds to a boosted Schwarzschild solution; performing this
boost, we can without loss of generality assume that $f_{Schw}=1$. The
analogous assumption will be made in the cases with
$\Lambda\not=0$ in the following.)
Some of the functions $f_{i,j}$ may vanish, but Chru\'sciel and
Singleton \cite{ChruSin} prove that there exist
space-times for which $f_{15,1}$ is non-vanishing.
This implies a surprising fact that, although there exist
extensions through the null hypersurface ${\cal H}^+$ given by
$u=+\infty$ which are $C^{117}$, in general the Robinson-Trautman
metrics {\it cannot be extended smoothly}.
Also, there exist an infinite number of $C^5$ extensions
through ${\cal H}^+$. In particular, we may join
the radiative metrics to the Schwarzschild metric so that the
Robinson-Trautman space-time ``settles down''
to the Schwarzschild space-time including the interior of the
black hole, as shown in Fig.1.
In order to see the smoothness across ${\cal H}^+$, one
introduces an advanced time coordinate $v$ by
$v=u+2r^*=u+2r+4m\ln(r/2m-1)$,  and Kruskal-type coordinates
$\hat u, \hat v$ by (see, e.g., \cite{Tod})
\begin{eqnarray}
&&\hat u= -\exp(-u/4m) \ , \nonumber\\             \label{E2.14}
&&\hat v=\quad\exp( v/4m) \ .
\end{eqnarray}
The hypersurface $u=+\infty$ now becomes a boundary given by
$\hat u=0$. The metric (\ref{E2.2}) becomes
\begin{equation}
ds^2=-\frac{32m^3}{r}\exp(-r/2m) d\hat u d\hat v - 16m^2\hat\Phi
d\hat u^2 + 2r^2P^{-2}d\zeta d\bar\zeta\ ,      \label{E2.15}
\end{equation}
where
\begin{equation}
\hat\Phi=e^{u/2m}\left(\frac{1}{2}R-1+
    \frac{r}{12m}\Delta_g R\right) ,            \label{E2.16}
\end{equation}
with $R$ and $\Delta_g$ being given by (\ref{E2.7}) (for
$f=1\Rightarrow \hat\Phi=0$ it reduces to the Schwarzschild space-time in
standard Kruskal coordinates). In terms of
$\hat u$, the expansion (\ref{E2.13}) becomes
\begin{eqnarray}
f&=&1+f_{1,0}\hat u^8+f_{2,0}\hat u^{16}+\cdots+f_{14,0}\hat u^{112} \nonumber\\
 && -4mf_{15,1}(\ln|\hat u|)(\hat u)^{120}+f_{15,0}\hat u^{120}+\cdots
 \ .                                             \label{E2.17}
\end{eqnarray}
Due to the presence of the $\ln|\hat u|$ terms, the function $f$ is
not smooth at $\hat u=0$; indeed it is $C^{119}$ if
$f_{15,1}\not=0$. The full metric (\ref{E2.15}) is $C^{117}$ at
$\hat u=0$, since $\hat\Phi$  contains the additional factor
$e^{u/2m}\sim 1/\hat u^2$.

\section {The Robinson-Trautman space-times with $0<9\Lambda m^2<1$}

When a Robinson-Trautman space-time with $\Lambda=0$ is known,
it is straightforward to
generalize it to the case of a non-vanishing $\Lambda$
(cf. \cite{KSMH, BiPo}). The metric still keeps the form
(\ref{E2.2}) with $P$ satisfying the equation (\ref{E2.4}).
The only place where $\Lambda$ enters is through the function $\Phi$.
The cosmological Robinson-Trautman metric reads
\begin{equation}
ds^2=-\Phi_\Lambda du^2 - 2dudr + 2r^2P^{-2}d\zeta d\bar\zeta\ ,\label{E3.1}
\end{equation}
where
\begin{equation}
\Phi_\Lambda=\Delta\ln P -2r(\ln P)_{,u} -
             \frac{2m}{r}-\frac{\Lambda}{3}r^2\ .  \label{E3.1a}
\end{equation}
We may still write $P=f\,P_0$, as in (\ref{E2.10}), where
$P_0$ gives (\ref{E2.9}) and $f$ satisfies (\ref{E2.6})-(\ref{E2.7}).
Since $\Lambda$ does not enter the
equation for $f$, we may take over the results for $\Lambda=0$
described in Section 2.  Therefore, as $u\rightarrow\infty$,
the metric (\ref{E3.1}) will now
approach the Schwarzschild-de Sitter metric given by $f=1$
corresponding to $\Phi_\Lambda^0=1-2m/r-\Lambda r^2/3$,
\begin{eqnarray}
ds^2&=&-\left(1-\frac{2m}{r}-\frac{\Lambda}{3}r^2\right)du^2-
  2dudr+r^2(d\theta^2+\sin^2\theta d\varphi^2)   \label{E3.1b} \\
    &=&-\left(1-\frac{2m}{r}-\frac{\Lambda}{3}r^2\right)dt^2+
      \left(1-\frac{2m}{r}-\frac{\Lambda}{3}r^2\right)^{-1}dr^2+
      r^2(d\theta^2+\sin^2\theta d\varphi^2) \ . \nonumber
\end{eqnarray}
Again, $u=t-r^*$, but the ``tortoise-type'' coordinate
$r^*$ for $0<9\Lambda m^2<1$ is
\begin{equation}
r^*=\int\frac{dr}{\Phi_\Lambda^0(r)}=
\delta_+\ln\frac{|r-r_+|}{r+r_++r_{++}}-
\delta_{++}\ln\frac{|r_{++}-r|}{r+r_++r_{++}}+
\delta_+\left[\ln\left(\frac{r_{++}}{r_+}\right)-\frac{1}{2}\right]\ ,
                                                  \label{E3.1c}
\end{equation}
where
\begin{equation}
\delta_{+}=\frac{r_+}{1-\Lambda r_+^2}\ \ ,\ \
\delta_{++}=-\frac{r_{++}}{1-\Lambda r_{++}^2}\ .
                                                  \label{E3.2d}
\end{equation}
Here $r_+=(2/\sqrt\Lambda)\cos(\alpha/3+4\pi/3)$,
with $\cos\alpha=-3m\sqrt\Lambda$, describes the black-hole horizon, and
$r_{++}=(2/\sqrt\Lambda)\cos(\alpha/3)$ is the
cosmological horizon --- see, e.g., \cite{BiPo} for more details
about dependence of parameters on $\Lambda$. (Analytic continuation
of the Schwarzschild-de Sitter metric is
discussed, for example, in \cite{Lake} and \cite{LW}-\cite{CuLa}.)

The presence of a cosmological constant does not affect the smoothness
of future infinity ${\cal I}^+$ in these space-times; however,
${\cal I}^+$ becomes spacelike for $\Lambda>0$ in
contrast to the cases with $\Lambda=0$ (cf. Fig.2).
Moreover, the presence of $\Lambda$ has a considerable effect
on the smoothness of extensions through
${\cal H}^+$ given by $u=+\infty$. The approach of $f$ to its
Schwarzschild-de Sitter form $f=1$ is again characterized
by the expansion (\ref{E2.13}) but
the transformation to Kruskal-type coordinates is
now given by
\begin{eqnarray}
&&\hat u=-\exp(-u/2\delta_+)\ , \nonumber\\             \label{E3.2e}
&&\hat v=\quad\exp( v/2\delta_+) \ ,
\end{eqnarray}
where $v=u+2r^*$, $r^*$ being given by (\ref{E3.1c}).
Hence, instead of (\ref{E2.17}), we get the expansion
\begin{eqnarray}
f&=&1+f_{1,0}(-\hat u)^{4\delta_+/m}+f_{2,0}(-\hat u)^{8\delta_+/m}+\cdots+
 f_{14,0}(-\hat u)^{56\delta_+/m} \nonumber\\
 && -2\delta_+ f_{15,1}(\ln|\hat u|)(-\hat u)^{60\delta_+/m}+
 f_{15,0}(-\hat u)^{60\delta_+/m}+\cdots
                                                \label{E3.3}
\end{eqnarray}
at $u\rightarrow+\infty$, i.e. $\hat u\rightarrow0_-$
(cf.(\ref{E3.2e})). The full metric takes the form
\begin{eqnarray}
ds^2&=&-\frac{4\Lambda\delta_+^2 r_+ e^{\frac{1}{2}}}{3r_{++}r}
  (r_{++}-r)^{1+\delta_{++}/\delta_+}(r+r_++r_{++})^{2-\delta_{++}/\delta_+}
  d\hat u d\hat v  \nonumber \\
 &&-4\delta_+^2\hat\Phi_\Lambda\,d\hat u^2
 +2r^2P^{-2}d\zeta d\bar\zeta \ ,              \label{E3.4}
\end{eqnarray}
where
\begin{equation}
\hat\Phi_\Lambda=e^{u/\delta_+}\left(\frac{1}{2}R-1+
    \frac{r}{12m}\Delta_g R\right) ,            \label{E3.5}
\end{equation}
with $f$ being of the form (\ref{E3.3}) above. We may join the
radiative Robinson-Trautman metrics with $\Lambda>0$ to the
Schwarzschild-de Sitter metric so that the space-time
 ``settles down'' to the Schwarzschild-de Sitter black-hole
including its interior (see Fig.2). Such
an extension across $\hat u=0$ will, in general, be $C^{5}$ in the case
of vanishing $\Lambda$. (For example, $\hat\Phi$ and all its
derivatives vanish for $\hat u=0$ in the Schwarzschild case,
whereas $\partial_{\hat u}^{(6)}\hat\Phi\not=0$ with $f$ given by
equation (\ref{E2.17}).)
With $\Lambda>0$, much higher smoothness can be
obtained. For those values of $\Lambda$ which imply $4\delta_+/m$
equals an integer, the smoothness is always better than for $\Lambda=0$.
Moreover, the horizon ${\cal H}^+$ can be made
``arbitrarily smooth'' by letting $\Lambda$ approach
its extremal value, $\Lambda\rightarrow 1/9m^2$
(i.e., $r_+\rightarrow 3m$). Then $\delta_+$  becomes arbitrarily
large  and the terms $\ \sim(-\hat u)^{i\delta_+/m}$, $i=4, 8,
\cdots$ in (\ref{E3.3})
will guarantee arbitrarily high smoothness of the function $f$ at
$\hat u=0$.

The Robinson-Trautman metrics with $\Lambda>0$ may serve
as exact analytic models  demonstrating the cosmic no-hair
conjecture under the presence of gravitational waves ---
they all approach de Sitter space-time locally
close to ${\cal I}^+$, i.e., near $r\rightarrow\infty$, $u$ finite
(cf. Fig.2). As discussed in detail in \cite{BiPo}, the transformation
of the form
\begin{eqnarray}
r     &=& \chi e^{H\tau}-H^{-2}(f_{\infty,u}/f_\infty)+\sum_{n=1}^\infty
  A_n\,e^{-nH\tau}\ ,   \nonumber \\
e^{Hu}&=&H\chi-e^{-H\tau}\quad+\   \sum_{n=3}^\infty
  B_n\,e^{-nH\tau}\ ,           \label{E3.6} \\
\zeta &=& \eta\quad+\qquad\qquad\ \  \sum_{n=3}^\infty
  C_n\,e^{-nH\tau}\ ,   \nonumber
\end{eqnarray}
in which $A_n$, $B_n$, $C_n$ are suitable functions of
$\chi,\eta,\bar\eta$, and $H=\sqrt{\Lambda/3}$,
 brings the metric (\ref{E3.1})
into the asymptotic form
\begin{eqnarray}
ds^2=&-d\tau^2+e^{2H\tau} \left[ d\chi^2+
      f_\infty^{-2} \chi^2\,(d\theta^2+\sin^2\theta d\varphi^2)
      \right]              \nonumber \\
&+\sum_{m=0}^\infty e^{-mH\tau}  h_{ab}^{(m)}\, dx^a dx^b\ ,
                               \label{E3.7}
\end{eqnarray}
where  the coordinates $\theta$, $\phi$ are reintroduced by
$\eta=\sqrt2 e^{i\varphi}\tan{(\theta/2)}$,
$f_\infty=f|_{\tau\rightarrow\infty} = f(u=
H^{-1}\ln|H\chi|,\theta,\phi)$, and $h_{ab}^{(m)}$ depend on
$\{x^a\} = \{\chi,\theta,\phi\}$ only.
It is seen explicitly that for $\tau\rightarrow\infty$ the
metric (\ref{E3.7}) does  not approach the de Sitter metric
globally --- the gravitational waves leave ``an imprint'' on
${\cal I}^+$ which is demonstrated by the presence of the function
$f_\infty$. However, any geodesic observer  will see
{\it locally}, inside his past light cone, space-time approach
de Sitter space-time exponentially fast in accordance with the
cosmic no-hair conjecture
(see \cite{BiPo} for details).

\section {The Robinson-Trautman space-times with $9\Lambda m^2=1$}

Above we summarized the approach to Schwarzschild-de Sitter
space-time in the case $0<9\Lambda m^2<1$ characterized by the
existence of two distinct horizons $r_+$ and $r_{++}$, with
$0<2m<r_+<3m<r_{++}$. With $\Lambda$ approaching its extremal value,
$\Lambda\rightarrow1/9m^2$, the black-hole horizon $r_+$
monotonically increases and the cosmological horizon $r_{++}$
decreases to the common value $3m$. In this section we shall analyze
the extreme case $9\Lambda m^2=1$ for which there exists
only one `double' Killing horizon at $r_e=3m$.

The metric of the  Robinson-Trautman space-time is still given by
(\ref{E3.1})-(\ref{E3.1a}), and the corresponding extreme Schwarzschild-de
Sitter metric  by (\ref{E3.1b}). However, the
``tortoise-type'' coordinate $r^*$ is now
\begin{equation}
r^*=\frac{9m^2}{r-3m}+2m\ln\left|\frac{r+6m}{r-3m}\right|\ ,
                                                  \label{E4.1}
\end{equation}
where an additive constant was chosen such that $r^*\rightarrow0$
at $r\rightarrow\infty$.
By introducing  the Kruskal-type null coordinates
\begin{eqnarray}
&&\hat u=-\arccot(-u/\delta)\ , \nonumber\\       \label{E4.2}
&&\hat v=\ \ \>\arctan( v/\delta) \ ,
\end{eqnarray}
where
\begin{equation}
\delta=-m(3-2\ln2)<0 \ ,                         \label{E4.2a}
\end{equation}
$v=u+2r^*$, $r^*$ given by
(\ref{E4.1}), the `extreme' Robinson-Trautman metric can be written
in the form
\begin{equation}
ds^2=-\frac{\delta^2}{27m^2 r}
  \frac{(r+6m)(r-3m)^2}{\cos^2\hat v \sin^2\hat u}d\hat u d\hat v
 -\hat\Phi_\Lambda\,d\hat u^2  +2r^2P^{-2}d\zeta d\bar\zeta \ ,
                                                \label{E4.3}
\end{equation}
where
\begin{equation}
\hat\Phi_\Lambda=\frac{\delta^2}{\sin^4\hat u}\left(\frac{1}{2}R-1+
    \frac{r}{12m}\Delta_g R\right)\  .            \label{E4.4}
\end{equation}
The asymptotic expansion (\ref{E2.13}) becomes
\begin{eqnarray}
f= \sum_{i,j\ge0} f_{i,j} \delta^j \cot^j \hat u\>
     e^{-(2i\delta/m)\cot\hat u}=
1+f_{1,0}e^{-(2\delta/m)\cot\hat u}+f_{2,0}e^{-(4\delta/m)\cot\hat u}
  +\cdots       \nonumber\\
  +f_{14,0}e^{-(28\delta/m)\cot\hat u}
   +\delta f_{15,1}\cot\hat u\> e^{-(30\delta/m)\cot\hat u}+\cdots
 \ .                                            \label{E4.5}
\end{eqnarray}
In particular, if $f=1$ we get $\hat\Phi_\Lambda=0$, $P=P_0$ (see
Eqs. (\ref{E2.10}), (\ref{E2.7}), (\ref{E2.9})), and the metric (\ref{E4.3})
describes the spherically symmetric extreme Schwarzschild-de Sitter
space-time --- see Fig.3 for its conformal diagram.
It is regular on the horizon $r=r_e=3m$
for all finite $u$ and $v$ since
\begin{equation}
\lim_{r\rightarrow 3m}\frac{(r-3m)^2}{\cos^2\hat v}=
\lim_{r\rightarrow 3m}\frac{(r-3m)^2}{\sin^2\hat u}=
\frac{(18m^2)^2}{\delta^2}\ .                  \label{E4.6}
\end{equation}

As in the previous case, the general Robinson-Trautman
space-times with $9\Lambda m^2=1$ approach an extreme
Schwarzschild-de Sitter space-time as $u\rightarrow+\infty$, i.e.,
$\hat u\rightarrow 0_-$ ($\hat u<0$ --- cf. (\ref{E4.2})). Indeed, introducing
$g_i=a_i\cot\hat u$, where $a_i=-(2i\delta/m)>0$, $i=1,2,3,\cdots$, and
$h_j=\exp(g_j)$, $j=1,2,3,\cdots$, the expansion of the function $f-1$
given by (\ref{E4.5}) can be written as a linear combination of terms
$(g_i)^k\,h_j$, $k=0,1,2,\cdots$. Clearly, $g_i\rightarrow -\infty$ as
$\hat u\rightarrow 0_-$, so that $(g_i)^k\,h_j\rightarrow 0$; this
implies $f\rightarrow 1$. All Robinson-Trautman
spacetimes (\ref{E4.3})-(\ref{E4.5}) are thus settling down to
the extreme Schwarzschild-de Sitter space-time as
$u\rightarrow\infty$, i.e., at the null hypersurface
${\cal H}^+$  given by  $\hat u=0_-$  (see
Fig.4). A question again naturally arises, whether one can extend
the space-time through ${\cal H}^+$ by glueing  to them, for
example, an extreme Schwarzschild-de Sitter
space-time (with $\hat u>0$).
It is not difficult to see that one can make such an extension, and,
in contrast to the cases $0\le9\Lambda m^2<1$, this {\it extension is smooth}.

First, it can be shown by induction and using the
relation $dg_i/d\hat u=-(a_i+g_i^2/a_i)$ that the $n$-th derivative,
$n=1,2,\cdots$, of $(g_i)^k$ with respect to $\hat u$ can be expressed
as a polynomial of the ($n+k$)-th order in $g_i$, i.e.,
$(g_i^k)^{(n)}= \sum_{s=0}^{n+k} c_{ks} g_i^s$, where the
coefficients $c_{ks}$ are
constants. Similarly, $h_j^{(n)}= h\>\sum_{s=0}^{2n} d_s g_j^s$,
where $d_s$ are constants. Leibnitz's formula then gives
$(g_i^k h_j)^{(n)}\rightarrow 0$ as $g_i\rightarrow-\infty$ which implies
\begin{equation}
\lim_{\hat u\rightarrow 0_-}f      =1\ ,\qquad
\lim_{\hat u\rightarrow 0_-}f^{(n)}=0\ .   \label{E4.7}
\end{equation}
Moreover, we find
\begin{equation}
\lim_{\hat u\rightarrow 0_-}(\hat\Phi_\Lambda)^{(n)}=0\ ,\label{E4.8}
\end{equation}
since $\sin ^{-4}\hat u\sim g_i^4\,\cos ^{-4}\hat u$,
so that $\hat\Phi_\Lambda=\{{\rm linear\  combination\  of\ }
g_i^{k+4}h_j\}\, \cos^{-4}\hat u$; an arbitrary derivative of the
first factor tends to zero as $\hat u\rightarrow 0_-$ while
derivatives of the second factor remain finite.

 Therefore, the radiative Robinson-Trautman space-times
with $9\Lambda m^2=1$ can be extended {\it smoothly}
through the horizon ${\cal H}^+$ to
the spherically symmetric extreme Schwarz\-schild-de Sitter
space-time with the same  values of $\Lambda$ and $m$, $9\Lambda m^2=1$
(see Fig.4). However, such an extension is not unique.
There are other possibilities, the simplest one can be obtained
by glueing a copy of the Robinson-Trautman space-time with $9\Lambda m^2=1$
to itself (see Fig.5). For $\hat u>0$ we consider another copy
of (\ref{E4.3})-(\ref{E4.5}) obtained by the reflection
$\hat u\rightarrow-\hat u$, $\hat v\rightarrow-\hat v$.
(The same reflection connects Figs.3a and 3b.)
Again, since $\lim_{\hat u\rightarrow 0_+}(\hat\Phi_\Lambda)^{(n)}=0$, the
extension across $\hat u=0$ is smooth and the space-time can be
called an `extreme' Robinson-Trautman black hole in the de
Sitter universe. Its conformal diagram resembles the diagram in
Fig.2 representing the non-extreme case (cf. \cite{BiPo}). Any
time-like geodesic observer falling from the region $\hat u<0$
will cross the smooth horizon ${\cal H}^+$ and reach the
singularity at $r=0$, or escape to `de Sitter-like' infinity
given by $r=\infty$.

 Therefore, the smooth extensions across $\hat u=0$ are not
unique. Of course, {\it they are not analytic}. In fact, the
functions $\exp(a_i\cot\hat u)$ in  expansion (\ref{E4.5})
are $C^\infty$ at $\hat u=0_-$, but $\hat u=0$ is an
irremovable singularity.

The behaviour of the  Robinson-Trautman space-times near
future spacelike infinity ${\cal I}^+$ (given by $r=\infty$)
is similar to the non-extreme case
discussed in the previous section. Again, one can perform the
transformation (\ref{E3.6}) converting the metric into the
asymptotic form (\ref{E3.7}) so that those space-times approach
the de Sitter metric locally as $\tau\rightarrow\infty$, in
correspondence with the cosmic no-hair conjecture.

\section {The Robinson-Trautman space-times with $9\Lambda m^2>1$}

In this case the corresponding Schwarzschild-de Sitter space-time
(\ref{E3.1b}) admits no horizon in the region $r>0$
(cf. \cite{Lake, CuLa}) so that there is only a naked singularity
situated at $r=0$. The metric of the  Robinson-Trautman space-time
with $9\Lambda m^2>1$ is again given by (\ref{E3.1})-(\ref{E3.1a}) but now
the `tortoise-type' coordinate $r^*$ becomes
\begin{equation}
r^*=-\frac{r_-}{\Lambda r_-^2-1}\left\{
\frac{1}{2}\ln\frac{r^2-2r_-r+r_-^2}
{r^2+r_-r-\frac{6m}{\Lambda r_-}}
+\frac{\frac{r_-}{2}-\frac{6m}{\Lambda r_-^2}}
{\sqrt{\frac{3}{4}r_-^2-\frac{3}{\Lambda}}}
\left[ \arctan\left(\frac{r+\frac{r_-}{2}}
{\sqrt{\frac{3}{4}r_-^2-\frac{3}{\Lambda}}}\right)
-\frac{\pi}{2}\right]\right\}\  ,                      \label{E5.1}
\end{equation}
where $r_-=-(3m/\Lambda)^{1/3}[(1-C)^{1/3}+(1+C)^{1/3}]<0$ and
$C=\sqrt{1-1/(9\Lambda m^2)}$. It can be shown that
$r^*$ monotonically decreases from $r^*(r=0)>0$ to
$r^*(r=\infty)=0$.
The Kruskal-type  coordinates are
\begin{eqnarray}
&&\hat u=-\arccot(u/m)\ , \nonumber\\       \label{E5.2}
&&\hat v=\arctan(-v/m) \ ,
\end{eqnarray}
where $v=u+2r^*$ and $r^*$ is given by
(\ref{E5.1}). Then the  Robinson-Trautman metric reads
\begin{equation}
ds^2=-\frac{\Lambda m^2}{3r}
  \Big(r-r_-\Big)\Big(r^2+r_-r-\frac{6m}{\Lambda r_-}\Big)
  \frac{d\hat u d\hat v}{\sin^2\hat u \cos^2\hat v }
 -\hat\Phi_\Lambda\,d\hat u^2  +2r^2P^{-2}d\zeta d\bar\zeta \ ,
                                                \label{E5.3}
\end{equation}
where
\begin{equation}
\hat\Phi_\Lambda=\frac{m^2}{\sin^4\hat u}\left(\frac{1}{2}R-1+
    \frac{r}{12m}\Delta_g R\right)\  ,          \label{E5.4}
\end{equation}
and
\begin{eqnarray}
f= \sum_{i,j\ge0} f_{i,j} (-m)^j \cot^j \hat u\>
     e^{2i\cot\hat u}=
1+f_{1,0}e^{2\cot\hat u}+f_{2,0}e^{4\cot\hat u}
  +\cdots       \nonumber\\
  +f_{14,0}e^{28\cot\hat u}
   -m f_{15,1}\cot\hat u\> e^{30\cot\hat u}+\cdots
 \ .                                            \label{E5.5}
\end{eqnarray}
The metric is regular for all values $r>0$ and, in particular, it
describes spherically symmetric Schwarzschild-de Sitter
space-time with a naked
singularity if $f=1$ (i.e., $\hat\Phi_\Lambda=0$); its conformal
diagram is seen in Fig.6. Since the expansion (\ref{E5.5})  is
analogous to (\ref{E4.5}), we can take over the results
(\ref{E4.7})-(\ref{E4.8}) implying that any Robinson-Trautman
space-time with $9\Lambda m^2>1$ approaches smoothly the
corresponding  Schwarzschild-de Sitter space-time as
$u\rightarrow\infty$ ($\hat u\rightarrow0_-$). It contains no
horizon (contrary to the cases discussed in the previous
sections) so that the metric (\ref{E5.3})-(\ref{E5.5})
need not to be extended past $\hat u=0$; it is already
geodesically complete for $u>u_0$, as indicated in Fig.7. Also,
it can be put into the asymptotic form (\ref{E3.7}), again demonstrating
explicitly the cosmic `no-hair' conjecture under the presence of
gravitational waves.

\section {The Robinson-Trautman space-times with $\Lambda<0$}

We now complete the analysis of the Robinson-Trautman
vacuum space-times with $\Lambda$ with the case $\Lambda<0$.
The Schwarzschild-anti-de Sitter metric, which is again a spherically
symmetric Robinson-Trautman solution given by (\ref{E3.1})-(\ref{E3.1a})
with $f=1$ (or (\ref{E3.1b}) with $\Lambda<0$),
always admits a black-hole horizon at
$r_h=(-3m/\Lambda)^{1/3}[(C+1)^{1/3}-(C-1)^{1/3}]>0$, where
$C=\sqrt{1-1/(9\Lambda m^2)}$. The value of $r_h$ decreases from
$r_h=2m$ for $\Lambda=0$ to $r_h\rightarrow0$ as
$\Lambda\rightarrow-\infty$, as seen in Fig.8 (the expansion of $r_h$
for small $\Lambda<0$ is $r_h=2m+\frac{8}{3}m^3\Lambda+{\cal
O}(m^5\Lambda^2)$). Kruskal-type null coordinates are
\begin{eqnarray}
&&\hat u=-\exp(-u/2\delta_h)\ , \nonumber\\
&&\hat v=\quad\exp( v/2\delta_h) \ .    \label{E6.1}
\end{eqnarray}
Here $v=u+2r^*$, with the `tortoise-type' coordinate $r^*$ for $\Lambda<0$
given by
\begin{eqnarray}
r^*&=&\delta_h\Bigg\{
 \ln|r-r_h|
-\frac{1}{2}\ln\Big(r^2+r_h r-\frac{6m}{\Lambda r_h}\Big)
                                         \label{E6.2}\\
&&+\frac{6m-r_h}{\sqrt{(6m+r_h)(2m-r_h)}}
\left[ \arctan\left(\sqrt{\frac{2m-r_h}{6m+r_h}}
  \Big(1+\frac{2r}{r_h}\Big)\right)
+D \right]   \Bigg\}\  ,     \nonumber
\end{eqnarray}
where
\begin{equation}
\delta_h=-\frac{3}{2\Lambda r_h}
    \frac{2m-r_h}{3m-r_h}\ ,              \label{E6.3}
\end{equation}
and $D=-m\sqrt{-\Lambda/3}\>[1+\ln(-4\Lambda m^2/3)]$. Performing
the transformation  (\ref{E6.1}), the Robinson-Trautman metric (\ref{E3.1})
becomes
\begin{eqnarray}
ds^2&=&\frac{4\Lambda\delta_h^2}{3r}
  \Big(r^2+r_h r-\frac{6m}{\Lambda r_h}\Big)^{3/2}\nonumber\\
&&  \exp\left\{-\frac{6m-r_h}{\sqrt{(6m+r_h)(2m-r_h)}}
\left[ \arctan\left(\sqrt{\frac{2m-r_h}{6m+r_h}}
  \Big(1+\frac{2r}{r_h}\Big)\right) +D \right] \right\}
  d\hat u d\hat v  \nonumber \\
 &&-4\delta_h^2\hat\Phi_\Lambda\,d\hat u^2
 +2r^2P^{-2}d\zeta d\bar\zeta \ ,              \label{E6.4}
\end{eqnarray}
where
\begin{equation}
\hat\Phi_\Lambda=e^{u/\delta_h}\left(\frac{1}{2}R-1+
    \frac{r}{12m}\Delta_g R\right)\  .            \label{E6.5}
\end{equation}
If $f=1$ we get $\hat\Phi_\Lambda=0$, and the metric reduces to the
Schwarzschild-anti-de Sitter metric in Kruskal coordinates;
its conformal diagram is indicated in Fig.9. Letting
$\Lambda\rightarrow0$, we obtain back the metric (\ref{E2.15}).
In a general case, the expansion (\ref{E2.13}) of $f$ in terms of
$\hat u$ introduced by (\ref{E6.1}) becomes
\begin{eqnarray}
f&=&1+f_{1,0}(-\hat u)^{4\delta_h/m}+f_{2,0}(-\hat u)^{8\delta_h/m}+\cdots+
 f_{14,0}(-\hat u)^{56\delta_h/m} \nonumber\\
 && -2\delta_h f_{15,1}(\ln|\hat u|)(-\hat u)^{60\delta_h/m}+
 f_{15,0}(-\hat u)^{60\delta_h/m}+\cdots \ .
                                                \label{E6.6}
\end{eqnarray}
Therefore, all radiative Robinson-Trautman metrics with
$\Lambda<0$ ``settle down'' to the Schwarzschild-anti-de Sitter
metric as $u\rightarrow\infty$, or $\hat u\rightarrow0_-$
(see Fig.10). However, the smoothness of the extension of the
Robinson-Trautman metric across the horizon ${\cal H}^+$ given by
$\hat u=0$ to the Schwarzschild-anti-de Sitter metric {\it
decreases} with a growing value of ($-\Lambda$). Indeed, the
parameter $\delta_h$ given by  (\ref{E6.3}) monotonically decreases from
$\delta_h=2m$ for $\Lambda=0$ to $\delta_h\rightarrow0$ as
$\Lambda\rightarrow-\infty$ (see Fig.8). For $(-\Lambda)$
small one gets $7<4\delta_h/m<8$, so that the function $f$ is at
least $C^7$ and the full metric is $C^5$ (the smoothness of the
extension is decreased by 2 due to the factor
$e^{u/\delta_h}\sim 1/\hat u^2$ entering $\hat\Phi_\Lambda$), as
in the case with $\Lambda=0$.
For $-\Lambda m^2>4/9$ the black-hole horizon is situated at
$r_h<3m/2$ and $4\delta_h/m<3$; the function $f$ is less then
$C^3$ and the  metric is not even $C^1$.
If $-\Lambda m^2>3$ then $r_h<m$, $4\delta_h/m<1$, and
$df/d\hat u$ diverges at ${\cal H}^+$.

However, as expected, the presence of a negative cosmological
constant does not affect the smoothness of  infinity
${\cal I}$ (although it changes its character:  ${\cal I}$
becomes timelike). Introducing a
coordinate $l=r^{-1}$ and  a conformal factor $\Omega=l$ in
(\ref{E3.1})-(\ref{E3.1a}), one finds (cf. \cite{BiPo})
\begin{equation}
\Omega^2 ds^2=2dudl-l^2 \Phi_\Lambda du^2 +
                     2P^{-2}d\zeta d\bar\zeta\ , \label{E6.7}
\end{equation}
where
\begin{equation}
\Phi_\Lambda=\Delta\ln P -2l^{-1}(\ln P)_{,u} -
             2ml-\frac{\Lambda}{3}l^{-2}\ .     \label{E6.8}
\end{equation}
It is easy to see that $l=0$ is a regular timelike hypersurface
for arbitrary smooth $P(u,\zeta,\bar\zeta)$.

\section {Concluding remarks}

We have shown that all vacuum radiative cosmological Robinson-Trautman
space-times of the Petrov type II with $m>0$ settle down to
Schwarzschild-de Sitter (if $\Lambda>0$) or
Schwarzschild-anti-de Sitter (if $\Lambda<0$) solutions at large
retarded times. This is true for ``arbitrary strong'' smooth
initial data in the Robinson-Trautman class of metrics. The
space-times can then be extended to include the black-hole
interiors. As $\Lambda>0$ is increased, the interior of a
corresponding Schwarzschild-de Sitter black
hole can be joined to an external cosmological
Robinson-Trautman space-time across the horizon with an increased
degree of smoothness. In the extreme case when $9\Lambda m^2=1$
the extension is $C^\infty$, i.e. smooth, but not analytic. In
this sense, the Conjecture 2.1 presented for the case $\Lambda=0$ in
Ref. \cite{Chru2}, that the only ``positive mass Robinson-Trautman
space-time which is smoothly extendible through ${\cal H}^+$ is
(necessarily) the Schwarzschild space-time'' is not true for
Robinson-Trautman space-times with a positive cosmological constant.
On the other hand, for $\Lambda<0$  the extension to
a Schwarzschild-anti-de Sitter black hole has a lower degree of
smoothness than in corresponding cases  with $\Lambda=0$.

All space-times with $\Lambda>0$
represent  exact explicit models exhibiting the cosmic no-hair
conjecture under the presence of gravitational waves.
They may serve as test beds in numerical studies of more realistic
situations.

\section {Acknowledgments}

JP thanks prof. D. Kramer for a kind hospitality at the F.
Schiller University in Jena where part of this work was done, and
JB acknowledges the hospitality of the M. Planck Institute for
Gravitational Physics in Potsdam where this work was completed.
We also acknowledge the support of Grants No. GACR-202/96/0206 and
No. GAUK-230/1996 from the Czech Republic and Charles University,
and US-Czech Science and Technology grant No. 92067.

\newpage
{\LARGE Figure Captions}

\bigskip
\bigskip
\noindent
Figure 1. Starting with arbitrary, smooth initial data at $u=u_0$,
the radiative Robinson-Trautman metrics with $\Lambda=0$ converge
exponentially fast to a Schwarzschild metric as $u\rightarrow\infty$.
However, extension beyond the null hypersurface
${\cal H}^+(u=+\infty)$ can only be done with a finite degree
of smoothness.

\bigskip
\bigskip
\noindent
Figure 2. Starting with initial data at $u=u_0$, the
Robinson-Trautman metrics with $0<9\Lambda m^2<1$ converge
to a Schwarzschild-de Sitter metric as $u\rightarrow\infty$.
Although traces of gravitational waves will persist at future
infinity ${\cal I}^+$ for all geodesic observers the metric will
approach the de Sitter metric within their past light cone.
The metric at the horizon ${\cal H}^+$ has only
a finite degree of smoothness,  although this can
be higher than the case with $\Lambda=0$.

\bigskip
\bigskip
\noindent
Figure 3. a) Conformal diagram of the extreme Schwarzschild-de
Sitter space-time with $9\Lambda m^2=1$ and the singularity in
the past, corresponding to a white hole. The maximal analytic
extension of the geometry is obtained by glueing an infinite
number of regions shown in the figure, or joining a finite number of
regions via identification of events along two horizons $r=3m$.
b) The time-reversed diagram ($\hat u\rightarrow-\hat u, \hat v
\rightarrow-\hat v$), corresponding to a black hole.

\bigskip
\bigskip
\noindent
Figure 4. Starting with initial data at $u=u_0$, the
Robinson-Trautman metrics with $9\Lambda m^2=1$ converge
to an extreme Schwarzschild-de Sitter space-time as
$u\rightarrow\infty$. The extension beyond the horizon
${\cal H}^+$ is smooth but not analytic.

\bigskip
\bigskip
\noindent
Figure 5. Another smooth extension of the Robinson-Trautman
metric with $9\Lambda m^2=1$ beyond the horizon
${\cal H}^+$ can be obtained by glueing two copies of the metric
along $u=\infty$ ($\hat u=0$). The extreme black-hole space-time
illustrated in Fig.3b can also be joined to the Robinson-Trautman
space-time along ${\cal H}^+$.

\bigskip
\bigskip
\noindent
Figure 6. Conformal diagram of the Schwarzschild-de Sitter
space-time with $9\Lambda m^2>1$ describing a spherically
symmetric naked singularity in the (asymptotically) de Sitter universe.

\bigskip
\bigskip
\noindent
Figure 7. Starting with smooth initial data at $u=u_0$, the
Robinson-Trautman metrics with $9\Lambda m^2>1$ approach
a `naked' Schwarzschild-de Sitter metric as $u\rightarrow\infty$.
No extension of the metric is necessary.

\bigskip
\bigskip
\noindent
Figure 8. A plot of the black-hole horizion $r_h$ and the parameter
$\delta_h$ (dashed line) as a function of $\Lambda<0$ and $m$.

\bigskip
\bigskip
\noindent
Figure 9. Conformal diagram of the Schwarzschild-anti-de Sitter
space-time with $\Lambda<0$ and $m>0$. Infinity ${\cal I}$ is timelike.

\bigskip
\bigskip
\noindent
Figure 10. Starting with smooth initial data at $u=u_0$, the
Robinson-Trautman metrics with $\Lambda<0$ converge
to a Schwarzschild-anti-de Sitter metric as $u\rightarrow\infty$.
The metric at the horizon ${\cal H}^+$ has only
a finite degree of smoothness which is lower than in the case
with $\Lambda=0$.

\end{document}